%% file: main.tex
\PassOptionsToPackage{dvipsnames}{xcolor} %
\documentclass[sigconf]{acmart}

\AtBeginDocument{%
  \providecommand\BibTeX{{%
    \normalfont B\kern-0.5em{\scshape i\kern-0.25em b}\kern-0.8em\TeX}}}

\setcopyright{none}
\renewcommand\footnotetextcopyrightpermission[1]{} %

\include{packages}

\usetikzlibrary{tikzmark}    
\usetikzmarklibrary{listings}
\usepackage{subcaption}
\usepackage{multirow}

\title{Automatic Detection of Speculative Execution Combinations}

\author{Xaver Fabian}
\affiliation{%
  \institution{Cispa Helmholtz Center for Information Security}
  \city{Saarbr\"ucken}
  \country{Germany}
}
\email{xaver.fabian@cispa.de}

\author{Marco Guarnieri}
\affiliation{%
  \institution{IMDEA Software Institute}
  \city{Madrid}
  \country{Spain}
}
\email{marco.guarnieri@imdea.org}

\author{Marco Patrignani}
\affiliation{%
  \institution{University of Trento}
  \city{Trento}
  \country{Italy}}
\email{marco.patrignani@unitn.it}

\begin{document}

\begin{abstract}
Modern processors employ different prediction mechanisms to speculate over different kinds of instructions.
Attackers can exploit these prediction mechanisms \emph{simultaneously} in order to trigger {leaks} about speculatively-accessed data.
Thus, sound reasoning about such speculative leaks requires accounting for \emph{all} potential mechanisms of speculation.
Unfortunately, existing formal models only support reasoning about fixed, hard-coded mechanisms of speculation, with no simple support to extend said reasoning to new mechanisms.

In this paper we develop a framework for reasoning about composed speculative semantics that capture speculation due to different mechanisms and implement it as part of the \tool verification tool.
We implement novel semantics for speculating over store and return instructions and combine them with the semantics for speculating over branches.
Our framework yields speculative semantics for speculating over any combination of those instructions that are secure by construction, i.e., we obtain these security guarantees for free.
The implementation of our novel semantics in \tool let us verify existing codebases that are vulnerable to \specb, \specs, and \specr vulnerabilities as well as new snippets that are only vulnerable to their compositions.
\end{abstract}

\maketitle

\section{Introduction}\label{sec:intro}

{Speculative execution} avoids pipeline stalls by predicting intermediate results and by speculatively executing instructions based on such predictions. 
When a prediction turns out to be incorrect, the processor squashes the speculative instructions, thereby rolling back their effect on the architectural state. 
Speculative instructions, however, leave footprints in microarchitectural components (like caches) that persist even after speculative execution terminates.
As shown by Spectre~\cite{spectre}, attackers can exploit these side effects to leak information about speculatively accessed data.

Modern general-purpose processors have different prediction mechanisms (branch predictors, memory disambiguators, etc.) that are used to speculate over different kinds of instructions: conditional branching~\cite{spectre}, indirect jumps~\cite{spectre}, store and load operations~\cite{S_specv4}, and return instructions~\cite{spectreRsb}. 
While well-known attacks target only a single speculation mechanism (e.g., Spectre-PHT~\cite{spectre} targets branch predictors), some speculative leaks only arise due to the interaction of multiple speculation mechanisms.

\begin{lstlisting}[basicstyle=\small,style=Cstyle, 
    caption={Speculative leak arising from speculation over branch and store instructions combined.}, 
    label=lst:v1-v4-combined,escapechar=|, captionpos=t]
x = 0; |\label{line:v14as}|
p = &secret; 
p = &public;    |\label{line:v14pub}|
if (x != 0)     |\label{line:v14branch}| 
    temp &= A[*p];      |\label{line:v14leak}|
\end{lstlisting}
For example, the code in \Cref{lst:v1-v4-combined} can speculatively leak the value of \inlineCcode{&secret} in \Cref{line:v14leak} whenever (1) the memory write to \inlineCcode{p} in \Cref{line:v14pub} is predicted to have a different address then the memory read \inlineCcode{*p} on \Cref{line:v14leak}, and (2) the branch instruction on \Cref{line:v14branch} is mispredicted as taken.
This leak, therefore, arises from the \emph{combination} of speculation over the branch predictor and the memory disambiguator.
Hence, leaks like the one in \Cref{lst:v1-v4-combined} are missed by \emph{sound} analyses for speculative leaks that consider speculation over \emph{only one} of these speculation mechanisms. %

Sound reasoning about speculative leaks requires accounting for \emph{all} potential sources of speculative execution.
However, existing formal models (also called \emph{speculative semantics}) support multiple sources of speculation poorly.
Some of them support only fixed speculation mechanisms: branch predictors~\cite{spectector,ST_spectector2, oo7,kleeSpectre,ST_blade} and (in addition) memory disambiguators~\cite{ST_constantTime_Spec, ST_binsec, cats}.
Furthermore, the different speculation mechanisms are \emph{hard-coded} into the formal semantics~\cite{spectector,ST_constantTime_Spec, ST_binsec}.
Extending these semantics with new speculation mechanisms (e.g., speculation over return addresses or value prediction) requires changes to the formal model and to any security proof relying on it.
This is simply not a scalable approach for developing comprehensive formal models and analyses for speculative leaks.\looseness=-1

In this paper we develop a framework for composing speculative semantics that capture speculation due to different mechanisms and implement it as part of the \tool verification tool.
The combination yields a single operational semantics which can be used to reason about leaks involving the different kinds of speculation that compose it (as in \Cref{lst:v1-v4-combined}).
Our framework lets us define the speculative semantics of each mechanism independently, which leads to \emph{simpler formalisation}.
Additionally, the security of the composed semantics can be derived automatically from the security of its sub-parts, maximising \emph{proof reuse}.
Finally, the composed semantics can be easily implemented in \tool, which can then be used to \emph{verify the absence of speculative leaks} such as those of \Cref{lst:v1-v4-combined}.

Concretely, this paper makes the following contributions:
\begin{itemize}
    \item it introduces $\sems$ and $\semr$, two novel semantics for speculation over store and return instructions (\Cref{sec:newsems}).
    
    \item it defines the framework for composing different speculative semantics and formalises its key properties: if the individual semantics fulfil some (expected) security conditions (which we prove for all the semantics we combine), then the composed semantics is also secure (\Cref{sec:frame}).

    \item it instantiates the framework with $\sems$, $\semr$ and $\semb$, the semantics for speculation over branch instructions~\cite{spectector}, creating all the possible compositions ($\sembs$, $\semsr$, $\sembr$, and $\sembsr$) and proving their security (\Cref{sec:comb-in}).
    
    All the presented semantics are formalised in the Coq Proof assistant, and we write \coqed to indicate that the semantics of some specific snippet is calculated mechanically.
        
    \item it extends the \tool verification tool with all these semantics and validates this extension on both existing benchmarks (for speculation on store and return instructions) as well as on new snippets (for combined speculation) that we define (\Cref{sec:impl-b}).
\end{itemize}
The rest of the paper first presents background notions, such as the security notion we rely on, and the formal language we extend with the novel speculative semantics (\Cref{sec:bg}) and then related work (\Cref{sec:rw}) and conclusions (\Cref{sec:conc}).

For space constraints, formal details of the semantics, auxiliary lemmas and proofs can be found in the companion technical report.
Our code extension to \tool will be open sourced.

\input{src/main}

\input{src/spec_semantics}

\input{src/composing_semantics}

\input{src/implementation}

\input{src/limitations}
\input{src/related_work}

\input{src/future_work}
\newpage
\bibliographystyle{ACM-Reference-Format}
\bibliography{main}

\appendix

\end{document}

%% file: packages.tex
\usepackage[utf8]{inputenc}

\usepackage{amsfonts}
\usepackage{amsmath}
\usepackage{halloweenmath}
\usepackage{color}
\usepackage{todonotes}
\usepackage{amsthm}

\usepackage{pifont}%
\usepackage{mathtools}
\usepackage{mathpartir}
\usepackage{natbib} %
\usepackage{relsize}
\usepackage{listings}
\usepackage{listingsutf8}
\usepackage{caption}
\usepackage{thmtools, thm-restate} 
\usepackage{booktabs}
\usepackage{paralist}

\usepackage{cleveref} %
\declaretheorem{theorem}

\declaretheorem[]{definition}

\theoremstyle{definition}

\newtheorem{example}{Example}

\theoremstyle{plain}

\usepackage{tikz}
\usetikzlibrary{positioning,shadows,arrows,calc,backgrounds,fit,shapes,snakes,shapes.multipart,decorations.pathreplacing,shapes.misc,patterns}
\usepackage{forest}
\usepackage{xspace}
\usepackage{theoremref} %

\usepackage{stmaryrd} %

\usepackage{xargs}
\usepackage{hyperref}

\usepackage{wrapfig}

\usepackage{soulutf8}
\newcommand{\ulc}[2]{\setulcolor{#1}\ul{#2}}

\colorlet{pred}{Maroon}
\colorlet{pgrn}{SeaGreen}%
\colorlet{pblu}{Cerulean}
\colorlet{porg}{BurntOrange}
\setul{}{1pt}

\usepackage[inference]{semantic}
\usepackage{semanticMacros}
\newcounter{typerule}
\crefname{typerule}{rule}{rules}

\newcommand{\typeruleInt}[5]{%
	\def\thetyperule{#1}%
	\refstepcounter{typerule}%
	\label{tr:#4}%
  \ensuremath{\begin{array}{c}#5 \inference{#2}{#3}\end{array}} 
}
\newcommand{\typerule}[4]{%
  \typeruleInt{#1}{#2}{#3}{#4}{\textsf{\scriptsize ({#1})} \\      }
}

\makeatletter
\xdef\@thefnmark{\@empty}

\makeatother

%% file: src/main.tex
\section{Background:\texorpdfstring{\muasm{}}{uASM}, Speculative Semantics and Security Definition }\label{sec:bg}
This section first describes the attacker model and the security definition we consider (\Cref{sec:attacker-model}).
Then it presents  the syntax (\Cref{sec:syntax-mu}) and the semantics (\Cref{sec:semantics-mu}) of \muasm{}, a simple assembly-style language, followed by $\semb$, the semantics for speculation over branch instructions (\Cref{sec:v1-semantics}).
Most of the notions that we overview next are taken from \citet{spectector}. %

\subsection{Attacker Model and Security Definition}\label{sec:attacker-model}\label{sec:sni}
We adopt a commonly-used attacker model \cite{spectector, ST_spectector2, ST_binsec, ST_constantTime_Spec, ST_inspectre, ST_jasmin2, S_sec_comp, ST_blade}: a passive attacker observing the execution of a program through events $\tau$.
These events, which we call \emph{observations}, model timing leaks through cache and control flow while abstracting away low-level microarchitectural details. %
\begin{align*}
    \Obs \bnfdef&\ \loadObs{n} \mid \storeObs{n} \mid \pcObs{n} \mid \callObs{f} \mid \retObs{n}
    &
    \tau \bnfdef&\  \empTr \mid \Obs
    \\
    \mid&\ 
     \startObsx{n} \mid \rollbackObsx{n} 
    &
        \tauStack \bnfdef&\ \nil \mid \tauStack\cdot \tau
\end{align*}
The $\storeObs{n}$ and $\loadObs{n}$ events denote read and write accesses to memory location $n$, so they model cache leakage.
In contrast, $\pcObs{}$, $\callObs{f}$, and $\retObs{n}$ events record the control-flow of the program. 
The $\startObsx{n}$ and $\rollbackObsx{n}$ observations denote the start and the finish of a \emph{speculative transaction}~\cite{spectector} (with identifier $n$) produced by the speculative semantics $x$ (we often use $x$ to range over any of the speculative semantics we define later).

An observation $\tau$ is either an event $\Obs$ or the empty observation $\empTr$. 
Traces $\tauStack$ are sequences of observations; we indicate sequences of elements $[e_1; \cdots; e_n]$ as $\bar{e}$, and adding an element $e$ to $\bar{e}$ as $\bar{e} \cdot e$.

The \textit{non-speculative projection} $\nspecProject{}$ \cite{spectector} of a trace $\tauStack$ deletes all speculative observations by removing all sub-traces enclosed between $\startObsx{n}$ and $\rollbackObsx{n}$. 
The remaining trace, then, captures all non-speculative observations. %

With this trace model we can define the security property we use in this paper: \emph{Speculative Non-Interference} (SNI)~\cite{spectector}. 
Intuitively, SNI requires that programs do not leak more information under the speculative semantics than under the non-speculative semantics.

SNI is parametric in a policy $\pol$ and in the speculative semantics $x$ it uses. 
The policy $\pol$ describes the public/low information of the program. 
We use the same policy $\pol$ as described by \citet{spectector}: a list describing public registers and public memory locations.
Two configurations $\sigma^1, \sigma^2$ are called \textit{low-equivalent} for a policy $\pol$, written $\sigma^1 \backsim_{\pol} \sigma^2$, if they agree on all register and memory locations in $\pol$. 
The \emph{speculative semantics} $x$ defines how the (speculative) traces describing the program behaviour are generated.
We indicate that program $p$ generates trace $\tauStack$ from state $\sigma$ with semantic $x$ as $\amTracevx{(p, \sigma)}{\tauStack}$.
We fix the maximal speculation window, i.e., the maximum number of speculative instructions, to a global constant $\omega$.
We formalise several speculative semantics in later sections. %

A program $p$ satisfies SNI (\Cref{def:sni}) for a speculative semantics $x$ iff \ulc{pred}{any pair of low-equivalent initial configurations $\sigma^1$ and $\sigma^2$} for program $p$ that \ulc{porg}{generate} the \ulc{pgrn}{same observations without speculation events}, then the two configurations \ulc{pblu}{generate the same observations considering speculation events too}.
\begin{definition}[SNI]\label{def:sni}
   $\snix \isdef~ \forall \sigma^1,\sigma^2$ 
    {if} \ulc{pred}{$\sigma^1 \backsim_{\pol} \sigma^2$} 
    {and} \ulc{porg}{$\amTracevx{(p, \sigma^1)}{\tauStack^1}$}
    {and} \ulc{porg}{$\amTracevx{(p, \sigma^2)}{\tauStack^2}$} 
    {and} \ulc{pgrn}{$\nspecProject{\tauStack^1} = \nspecProject{\tauStack^2}$} 
    {then} \expandafter\ulc{pblu}{$\tauStack^1 = \tauStack^2$}.
\end{definition}

\subsection{\muasm{}}\label{sec:syntax-mu}
\begin{gather*}
\begin{aligned}
\text{(Programs) } p  \coloneqq 
        &\ 
        n:i \mid p_1;p_2
    &
    \text{(Functions) } \mathcal{F} \coloneqq&\ \varnothing \mid \mathcal{F}; f \mapsto n 
    \\
    \text{(Registers) } x \in
        &\ 
        \Reg 
    &
    \text{(Values) }  n, l \in
        &\
        \Val = \Nat \cup \{\bot\}
\end{aligned}
    \\
\begin{aligned}
    \text{(Expressions) }  e \coloneqq
        &\ 
        n \mid x \mid \ominus e \mid e_1 \otimes e_2
    \\
    \text{(Instructions) } i \coloneqq
        &\ 
        \pskip \mid \passign{x}{e} \mid \pload{x}{e} \mid \pstore{x}{e} 
    \mid 
        \pjmp{e} 
    \\
    \mid 
    &\ 
        \pjz{x}{l} 
        \mid 
        \pcondassign{x}{e}{e'} \mid \pbarrier 
        \mid \pcall{f} \mid \pret
\end{aligned}
\end{gather*}
\muasm{} is an assembly-like language whose syntax is presented above.
\muasm{} programs $p$ are sequences of mappings from natural numbers $n$ (i.e., the instruction address) to instructions $i$ or $\bot$.
Instructions include skipping, register assignments, loads, store, indirect jumps, branching, conditional assignments, speculation barriers, calls, and returns.
Instructions can contain expressions and values.
The former come from the set $\Reg$, containing register identifiers and designated registers $\pc$ and $\spR$ modelling the program counter and stack pointer respectively, while the latter come from the set $\Val$, which includes natural numbers and $\bot$.

\subsection{Non-speculative Semantics of \muasm}\label{sec:semantics-mu}
\muasm{} has a small-step operational non-speculative semantics that describes how its programs execute.
This semantics judgement is $(p, \sigma) \nsarrow{\tau} (p, \sigma')$ and it reads: \emph{``a program state $\tup{p, \sigma}$ steps to a new program state $\tup{p, \sigma'}$ producing observation $\tau$''}.
Program states $\tup{p, \sigma}$ consist of the program $p$ and the configuration $\sigma$.
The program $p$ is used to look up the current instruction, whereas the configuration $\sigma$ is used to read from/write to the register file $a$ and the memory $m$. 
Memories map addresses (which are natural numbers) to values while registers map registers id to values.

Most of the rules of the semantics are standard and thus omitted, we present selected rules below.
The rules rely on the evaluation of expressions (indicated as $\exprEval{e}{a} = v$) where expression $e$ is evaluated to value $n$ under register file $a$.
In the rules,  $a[x \mapsto y]$, where $x \in \Reg \cup \Nat$ and $y\in\Val$, denotes the update of a map (memory or registers), whereas  $a(x)$ denotes reading from a map.
Finally, $\sigma(x)$, where $x \in \Reg$ and $\sigma = \tup{m,a}$, denotes $a(x)$. %
\begin{center}
\typerule{Store}
{
\select{p}{a(\pc)} = \pstore{x}{e} & n = \exprEval{e}{a}
}
{
(p, \tup{m, a}) \eval{p}{\storeObs{n}} (p, \tup{ m[n \mapsto a(x)], a[\pc \mapsto a(\pc)+1])}
}{ns-store}

\typerule{Beqz-Sat}
{
\select{p}{a(\pc)} = \pjz{x}{\lbl} &  a(x) = 0
}
{
(p, \tup{m, a}) \eval{p}{\pcObs{\lbl}} (p, \tup{ m, a[\pc \mapsto \lbl])}
}{ns-beqz-sat}

\typerule{Call}
{
    \instr{\pcall{f}} &  \bigf  \\
    a' = a[\pc \mapsto n, sp \mapsto a(sp) - 8] & m'= [a'(sp) \mapsto a(\pc) + 1] 
}
{	
    \tup{p, \tup{m, a}} \nsarrow{\callObs{f}} \tup{p, \tup{m', a'}}
}{t-callStack}

\typerule{Return}
{
    \instr{\pret} & l = m(a(sp)) 
    \\
    a' = a[pc \mapsto l, sp \mapsto a(sp) + 8]}
{
    \tup{p, \tup{m, a}} \nsarrow{\retObs{l}} \tup{p, \tup{m, a'}}
}{t-retStack}
\end{center}
Conditionals emit observations recording the outcome of the branch (\Cref{tr:ns-beqz-sat}), while memory operations emit observations recording the accessed memory (\Cref{tr:ns-store}).
A call to function $f$ is a jump to the function's starting line number $n$, as indicated by the function map $\mathcal{F}$. A call stores the return address on the stack at the value of the stack pointer $\spR$ and decreases $\spR$ (\Cref{tr:t-callStack}). A return does the inverse: it looks up the return address via the stack pointer $\spR$ and then increases the stack pointer (\Cref{tr:t-retStack}).

The \emph{non-speculative behaviour} $\behNs{p}$ of a program $p$ is the set of all traces generated from an initial state until termination using the reflexive-transitive-closure of the non-speculative semantics.

\subsubsection{Symbolic semantics}
Following~\cite{spectector}, we introduce a \emph{symbolic} non-speculative semantics $\nsarrowa{}$ that is at the basis of \tool{}'s analysis.
This symbolic semantics differs from $\nsarrow{}$ in two key ways: (1) concrete configurations $\sigma$ are replaced with symbolic configurations $\sigmaa$, and (2) path condition constraints are generated in the standard way and they are encoded as part of the symbolic trace $\tauStack$.
Given a symbolic trace $\tauStack$, $\conc{\tauStack}$ denotes the set of all concrete traces that can be obtained by concretising $\tauStack$ with values consistent with $\tauStack$'s path condition.
As before, we can introduce the \emph{symbolic non-speculative behavior} $\behNsa{p}$ of a program $p$, and $\conc{\behNsa{p}}$ is the set of all concrete traces derived from symbolically executing $p$.
As proved by~\citet{spectector}, $\behNs{p} = \conc{\behNsa{p}}$.

\subsection{\texorpdfstring{${\semb}$}{Spec-B}: Speculating Over Branch Instructions}\label{sec:v1-semantics}

To model and reason about the effects of speculation induced by the branch predictor, \citet{spectector} propose the three semantics: an always-mispredict semantics  (\Cref{sec:v1-am}), an oracle semantics (\Cref{sec:v1-oracle}), and a symbolic semantics (\Cref{sec:v1-symb}).
The always-mispredict semantics, our main focus, is a safe overapproximation of the oracle one, which depends on a prediction oracle that models the branch predictor.
Finally, the symbolic semantics, which is used in \tool{}, is the symbolic version of the always-mispredict semantics.
We summarize the properties of these three semantics in \Cref{sec:v1-prop}.
With a slight abuse of notation, we use $\semb$ to indicate both the three speculative semantics, and the AM one alone (since it is the most relevant one).

\subsubsection{Always-mispredict (AM) Semantics}\label{sec:v1-am}
At every branch instruction, the always-mispredict semantics first speculatively executes the wrong branch for a fixed number of steps and then continues with the correct one. 
As a result, this semantics is deterministic and agnostic to implementation details of the branch predictor~\cite{spectector}.

The state $\SigmaB$ of the AM semantics is stack of speculative instances $\PhiB$ where reductions happen only on top of the stack.
Whenever we start speculating, a new instance is pushed on top of the stack (\Cref{tr:v1-branch}).
The instance is then popped when speculation ends (\Cref{tr:v1-rollback}). 
Each instance $\PhiB$ contains the program $p$, a counter $\ctr$ that uniquely identifies the speculation instance, a configuration $\sigma$, and the remaining speculation window $n$ describing the number of instructions that can still be executed speculatively (or $\bot$ when no speculation is happening).
\begin{align*}
    \ti{Spec. States } \SigmaB  \bnfdef&\ \phiStackB
    &
    \ti{Spec. Instances } \PhiB \bnfdef&\ \tup{p, \ctr, \sigma, n}
\end{align*}
This judgement for the AM semantics is: $\SigmaB \specarrowB{\tau} \SigmaB'$.
\begin{center}
    \typerule{$\Bvr$:AM-branch}
    {\instr{\pjz{x}{\lbl}} & \conctauarrow{(p, \sigma)}{(p, \sigma')} & j = min(\omega, n)  \\
     \sigma'' = \sigma[\pc \mapsto l'] &  \tauStack = \tau \cdot \startObsB{\ctr} \cdot \pcObs{l}\\
     l' = {\begin{cases} \sigma´(\pc) + 1 & \text{if $\sigma'(\pc) = l$} \\
                        l & \text{if $\sigma'(\pc) \neq l$}
     \end{cases}
     }
    }
    {\tup{p, \ctr, \sigma, n + 1}
    \specarrowB{\tauStack}
    \tup{p, \ctr, \sigma', n} \cdot \tup{p, \ctr + 1, \sigma'', j}
    }{v1-branch}
    
    \typerule {$\Bvr$:AM-NoSpec}
    { p(\sigma(\pc)) \notin [{\pjz{x}{\lbl}; \fcolorbox{lightgray}{lightgray}{\Z}}] & \conctauarrow{\sigma}{\sigma'} 
    }
    {\tup{p, \ctr, \sigma, n + 1}
    \specarrowB{\tau}
    \tup{p, \ctr, \sigma', n}
     }{v1-noBranch}
     
    \typerule{$\Bvr$:AM-Rollback}
    { n' = 0\ \text{or}\ \text{p is stuck}  }
    { \tup{p, \ctr, \sigma, n} \cdot \tup{p, \ctr', \sigma', n'}
    \specarrowB{\rollbackObsB{\ctr}}
     \tup{p, \ctr', \sigma, n}
    }{v1-rollback}
\end{center}

As mentioned, \Cref{tr:v1-branch} pushes a new speculative state with the wrong branch, followed by the state with the correct one.
When speculation ends, \Cref{tr:v1-rollback} pops the related state.
All other instructions are handled by delegating back to the non-speculative semantics (\Cref{tr:v1-noBranch}). 
\Cref{tr:v1-noBranch} also contains a novel element, the parameter $\Z$ (in \fcolorbox{lightgray}{lightgray}{gray}), which indicates other instructions that are skipped. $\Z$ is part of our composition framework and we defer to \Cref{sec:expl-z} an in-depth explanation of its role. %

The always-mispredict behaviour $\behB{p}$ of a program $p$ is the set of all traces generated from an initial state until termination using the reflexive-transitive closure of $\specarrowB{\tau}$.

\subsubsection{Oracle Semantics}\label{sec:v1-oracle}
The oracle semantics explicitly models the branch predictor using an oracle $\oracB$ that relies on the branching history $h$ of the program $p$ to predict branch outcomes. %

Here, we quickly summarize the key differences with the AM semantics; see~\cite{spectector} for the full definition.
First, speculative instances are extended to track the branching history $h$.
Second, when executing a $\jzC{}$ instruction, the oracle is used to predict the branch outcome and to create a new speculative instance that is pushed on top of the stack. 
Finally, whenever the speculation window of an instance anywhere on the stack reaches $0$, the execution needs to be rolled back or committed. 
Thus, committing and rolling back speculations happen along the stack.
Rolling back deletes all the instances above the rolled back instance, while committing updates the configuration, the counter and the branching history $h$ of the instance below and the committed instance is deleted.

Similarly to before, the behaviour  $\SEbehB{p}$ of a program $p$ under the oracle semantics is the set of all traces generated from an initial state until termination. %

\subsubsection{Symbolic Semantics}\label{sec:v1-symb}
The symbolic speculative semantics $\semba$ works on symbolic speculative states $\SigmaBa$ and is used in the implementation of \tool~\cite{spectector}. 
The only two differences w.r.t. the AM semantics are that (1) concrete states $\SigmaB$ are replaced with symbolic states $\SigmaBa$, which store symbolic configurations $\sigmaa$ instead of concrete configurations $\sigma$, and (2) the semantics uses the symbolic non-speculative semantic instead of the concrete one.

The rules of the symbolic semantics look like those of the AM one and the behaviour $\behBa{p}$ of a program $p$ is defined as for the AM semantics.

\subsubsection{Properties of $\semb$}\label{sec:v1-prop}

\citet{spectector} prove several properties relating the three semantics we presented above, which were instrumental in proving \tool{}'s security.
We recap these properties in a single definition (\Cref{def:sss}), which we will prove for all semantics we present in this paper.
In the definition we indicate that a program $p$ satisfies \SNI{} w.r.t. the oracle semantics as $\SEsniB$.
\begin{definition}[Secure Speculative Semantics]\label{def:sss}
A semantics $\semx$ is secure (denoted \ssssem{\semx}) if:
\begin{itemize}
    \item Oracle Overapproximation:
    $\forall \orac\ldotp \SEsnix ~ \text{iff}~ \snix$
    \item Symbolic Consistency:
    $\behx{p} = \conc{\behxa{p}}$
    \item NS Consistency:
    $\nspecProject{\behx{p}} = \behNs{p} = \nspecProject{\SEbehx{p}}$
\end{itemize}
\end{definition}
Intuitively, a secure speculative semantics is made of three components, an AM, an oracle and a symbolic semantics.
First, the AM semantics must overapproximate the oracle semantics, so it is enough to check a program $p$ for \SNI{} w.r.t. the AM semantics~\cite[Theorem 1]{spectector}.
Then, since \tool uses the symbolic semantics in the implementation, the symbolic semantics must be faithful w.r.t. the AM one~\cite[Proposition 2]{spectector}.
Finally, both the AM and the Oracle semantics can recover the non-speculative behaviour of a program $p$ by applying the non-speculative projection on their traces~\cite[Propositions 1,3]{spectector}.
So we can execute $p$ only once to get the (non-)speculative behaviour of that program run.

\Cref{thm:v1-sss} states that $\semb$ is a secure speculative semantics.

\begin{theorem}[$\semb$ is \sss~\cite{spectector}]\label{thm:v1-sss}
$\ssssem{\semb}$
\end{theorem}

%% file: src/spec_semantics.tex
\section{Speculation on Stores and Returns}\label{sec:newsems}

This section defines $\sems$ and $\semr$, two novel speculative semantics that model the effects of speculative execution over store (\Cref{sec:v4-semantics}) and return instructions (\Cref{sec:v5-sem}). 
Similarly to $\semb$, for each speculation source we define three semantics: an always-mispredict semantics, an oracle semantics, and a symbolic semantics.
As before, we will mostly focus on the always-mispredict semantics, which safely over-approximates the oracle one, and we will use its symbolic version to reason about leaks using \tool{}.
Most formal details, as well as proofs, can be found in the companion technical report.

\subsection{\texorpdfstring{$\sems$}{Spec-S}: Speculation on Store Instructions}\label{sec:v4-semantics}

Modern processors write $\storeC$s to main memory asynchronously to reduce delays caused by the memory subsystem.
For this, processors employ a \textit{Store Queue} where not-yet-committed $\storeC$ instructions are stored before being permanently written to memory.
When executing a $\loadC{}$ instruction, the processor first inspects the store queue for a matching memory address.
If there is a match, the value is retrieved from the store queue (called \emph{store-to-load forwarding}), and otherwise the memory request is issued to the memory subsystem. 
To speed up computation, processors employ memory disambiguation predictors to predict if  memory addresses of loads and stores match. 
Since the prediction can be incorrect, processors may speculatively bypass a store instruction in the store queue leading to a $\loadC{}$ instruction retrieving a stale value.

\begin{example}[Store Speculation Vulnerability]
Consider the example in \Cref{lst:v4-vanilla}:

\begin{lstlisting}[basicstyle=\small,style=Cstyle, caption=Code vulnerable to store speculation., label=lst:v4-vanilla,escapechar=|, captionpos=t]
p = &secret; |\label{line:v4sec}|
p = &public; |\label{line:v4pub}|
temp = B[*p * 512]; |\label{line:leakv4}|
\end{lstlisting}
Assume that the $\storeC{}$ instructions in \Cref{line:v4sec} and \Cref{line:v4pub} are still in the \textit{store queue} and not yet committed to main memory.
A misprediction of the memory disambiguator for the $\loadC{}$ instruction in \Cref{line:leakv4} causes it to bypass the $\storeC{}$ instruction in \Cref{line:v4pub} and retrieve the value from the stale $\storeC{}$ instruction in \Cref{line:v4sec}. The speculative access of the memory is then leaked into the microarchitectural state by the array access into \inlineCcode{B}.
\end{example}

This section first introduces the extended trace model required to talk about speculation over store instructions (\Cref{sec:v4-trace}).
Next, it presents the speculative AM semantics (\Cref{sec:v4-am}) and the corresponding oracle semantics (\Cref{sec:v4-oracle}) and symbolic semantics (\Cref{sec:v4-symb}).
This semantics is a secure speculative semantics (\Cref{thm:v4-sss}).
\begin{theorem}[$\sems$ is \sss]\label{thm:v4-sss}
$\ssssem{\sems}$
\end{theorem}

\subsubsection{Extended Trace Model}\label{sec:v4-trace}

We extend the trace model $\Obs$ with $\startObsS{n}$ and $\rollbackObsS{n}$ observations to mark start and end of a speculative transaction $n$ started by a store bypass. Furthermore, we add a $\skipObs{n}$ observation which denotes that a $\storeC{}$ instruction at program counter $n$ was skipped.
\begin{align*}
    \ObsS \bnfdef&\ \Obs \mid \startObsS{n} \mid \rollbackObsS{n} \mid \skipObs{n}
\end{align*}

\subsubsection{Speculative Semantics}\label{sec:v4-am}
The overall structure of the $\sems$ semantics is similar to that of $\semb$: speculative execution is modeled using a stack of speculative states, instructions that do not start speculative transactions are executed by delegating back to the non-speculative semantics, and speculative transactions are rolled back whenever the speculative window reaches 0.
The key difference between $\sems$ and $\semb$ is the differing source of speculation: branches for $\semb$ and stores for $\sems$.

The states used for the speculative semantics of $\sems$ are similar to the states of $\semb$:
\begin{align*}
    \ti{Spec. States } \SigmaS  \bnfdef&\ \phiStackS
    &
    \ti{Spec. Instance } \PhiS \bnfdef&\ \tup{p, \ctr, \sigma, n}
\end{align*}

Judgement $\SigmaS \specarrowS{\tau} \SigmaS'$ describes how program state $\SigmaS$ steps to $\SigmaS'$ emitting observation $\tau$. Similar to $\semb$, reductions only happen on top of the stack.

\begin{center}
    \typerule{\Svr:AM-Store}
        {
            \instr{\pstore{x}{e}} & \conctauarrow{(p, \sigma)}{(p, \sigma')}  
            &  
            j = min(\omega, n) 
            \\
            \sigma'' = \sigma[\pc \mapsto \sigma(\pc) +1] 
            &
            \tau' = \tau \cdot \skipObs{\sigma(\pc)} \cdot \startObs{\ctr}
        }
        {
            \tup{p, \ctr, \sigma, n + 1}
        \specarrowS{\tau'}
        \tup{p, \ctr, \sigma', n}
            \cdot \tup{p, \ctr + 1, \sigma'', j}
        }{v4-skip-new}
\end{center}
To model the effect of bypassing a $\storeC{}$ instruction, \Cref{tr:v4-skip-new} skips the $\storeC{}$ instruction by increasing the program counter without updating the memory and starts a new speculative transaction by pushing a new speculative instance on top of the state. 
A $\loadC{}$ instruction loading from the same memory location as the skipped $\storeC{}$ instruction loads a stale value.

Similarly to $\semb$, all instructions that are not $\storeC{}$ instructions are handled by delegating back to the non-speculative semantics and when the speculation window reaches $0$, a roll back occurs that pops the topmost speculative instance from the stack.

The behaviour $\behS{p}$ is the set of all traces that are generated from an initial state until termination using %
the reflexive-transitive closure of $\specarrowS{\tau}$.\looseness=-1 %

\subsubsection{Oracle Semantics}\label{sec:v4-oracle}
Instead of skipping every $\storeC{}$ instruction speculatively, the oracle semantics employs an oracle $\orac$ that decides if the $\storeC{}$ instruction should be skipped or not. 
Similar to before, the behaviour $\SEbehS{p}$ of a program $p$ is the set of all traces starting from an initial state until termination using the reflexive-transitive closure of the oracle semantics.
    
\subsubsection{Symbolic Semantics}\label{sec:v4-symb}
Similarly to $\semba$, the symbolic speculative semantics $\semsa$ requires two changes w.r.t. the always-mispredict one:  concrete configurations $\sigma$ and the non-speculative semantics are replaced by symbolic configurations $\sigmaa$ and the symbolic non-speculative semantics respectively. 
The behaviour $\behSa{p}$ of a program $p$ is the set of all traces starting
from an initial state until termination using the reflexive-transitive
closure of the symbolic semantics.

\subsection{\texorpdfstring{$\semr$}{Spec-R}: Speculation on Return Instructions}\label{sec:v5-sem}

The return-stack-buffer (RSB) is a small stack used by the CPU to save return addresses upon $\pcall{}$ instructions. These saved return addresses are speculatively used when the function returns because that is faster than looking up the return address on the stack (stored in main memory).
This works well because return addresses rarely change during function execution. However, mispredictions lead to speculative execution which can be exploited by an attacker. 

\begin{example}[Return Speculation Vulnerability]
Consider the example in \Cref{lst:v5-example} and recall that register $\spR$ is used to find return addresses saved on the stack.
\begin{lstlisting}[style=MUASMstyle, caption={A program exploiting RSB speculation.}, label={lst:v5-example}, escapeinside=!!]
Manip_Stack:
    sp !$\xleftarrow{}$! sp + 8  !\label{v5:line1}!
    ret
Speculate:
    call Manip_Stack !\label{v5:line4}!
    load eax, secret !\label{v5:line5}!
    load edx, eax
    ret
Main:
    call Speculate
    skip
\end{lstlisting}
Each function call pushes a return address on the stack and decrements the $\spR$ register. 
After reaching the function \textit{Manip\textunderscore Stack} in \cref{v5:line1} the $\spR$ register is incremented. 
Thus, $\spR$ points to the previous return address on the stack, and the non-speculative execution continues in \textit{Main} and terminates.
However, because the return address of the call in \cref{v5:line4} is \cref{v5:line5} and is on top of the RSB, the CPU speculatively uses this return address and execution jumps to \cref{v5:line5}; thus, the secret is leaked. 
\end{example}

This section describes the AM semantics (\Cref{sec:v5-am}), the Oracle semantics (\Cref{sec:v5-oracle}), and the symbolic semantics (\Cref{sec:v5-symb}).
Then, it discusses formalising different implementations of the RSB in the CPU (\Cref{sec:v5-rsb}).
This semantics is a secure speculative semantics (\Cref{thm:v5-sss}).
\begin{theorem}[$\semr$ is \sss]\label{thm:v5-sss}
$\ssssem{\semr}$
\end{theorem}

\subsubsection{Speculative Semantics}\label{sec:v5-am}

Unlike before, the state of $\semr$ contains a model of the RSB which is used to retrieve return addresses instead of relying on the stack.
Thus, speculative instances of $\semr$ are extended with an additional entry $\Rsb$ for tracking the RSB, whose size is limited by a global constant $\Rsb_{size}$ denoting the maximal RSB size. 
A speculative instance $\PhiR$ now consists of the program $p$, the counter $\ctr$, the configuration $\sigma$, the speculation window $\omega$ and the RSB $\Rsb$. 
As before, a state $\SigmaR$ is a stack of speculative instances $\phiStackR$.
\begin{align*}
    \ti{Spec. States } \SigmaR  \bnfdef&\ \phiStackR
    &
    \ti{Spec. Instance } \PhiR \bnfdef&\ \tup{p, \ctr, \sigma, \Rsb, n}
\end{align*}
As before, the small-step operational semantics $\SigmaR \specarrowR{\tau} \SigmaR$, reductions happen at the top of the stack.
\begin{center}
\typerule{$\Rvr$:AM-Ret-Spec}
    {
        \instr{\pret} & 
        \conctauarrow{\sigma}{\sigma'}  
        \\
        \Rsb = \Rsb' \cdot l & 
        j = min(\omega, n) 
        &
        l \neq m(a(\spR))
        \\
        \sigma'' = \sigma[\pc \mapsto l, \spR \mapsto a(\spR) + 8] & 
        \tauStack = \tau \cdot \startObsR{\ctr} \cdot \retObs{l}
    }
    {\tup{p, \ctr, \sigma, \Rsb, n + 1}
    \specarrowR{\tauStack}
    \tup{p, \ctr, \sigma', \Rsb', n} \cdot \tup{p, \ctr + 1, \sigma'', \Rsb', j}
    }{v5-ret}
    
    \typerule{$\Rvr$:AM-Call}
    {
        \instr{\pcall{f}} &  
        \conctauarrow{\sigma}{\sigma'}
        \\
        \Rsb' = \Rsb \cdot (a(\pc) + 1) & 
        \vert \Rsb \vert < \Rsb_{size}
    }
    {
    \tup{p, \ctr, \sigma, \Rsb, n+ 1} 
    \specarrowR{\tau}
    \tup{p, \ctr, \sigma', \Rsb', n} 
    }{v5-call}
\end{center}

During $\pcall{}$ instructions (\Cref{tr:v5-call}), the return address is pushed on top of the RSB (if there is space available) and during $\pret$ instructions, the return address stored on the RSB is used if the entry on top of the RSB is different from the one stored on the stack (\Cref{tr:v5-ret}). 
Then, the rule creates a new speculative instance that uses the return address from the RSB $\Rsb$. 
Note that speculation only happens when the return address from the RSB differs from the one on the stack (stored in $m(a(\spR))$). 

Here, we overview how our semantics behaves with an empty/full RSB; full formalization is available in the technical report.
Intuitively, whenever the RSB is empty, executing a $\pret{}$ instruction does not cause speculation and we return to the address pointed by $\spR$.
In contrast, whenever the RSB is full, executing a $\pcall{}$ instruction does not add entries to the RSB, i.e., we model a so-called \emph{acyclic} RSB.%
\footnote{We follow the way AMD processors handle this kind of speculation~\cite{ret2spec}.}

The behaviour $\behR{p}$ is is the set of all traces generated from an initial state until termination using 
the reflexive-transitive closure of $\specarrowR{}$.

\subsubsection{Oracle Semantics}\label{sec:v5-oracle}
Unlike before, the oracle cannot decide the outcome of the $\pret$ instruction, because the CPU always uses the return address stored in the RSB (if there is one) and it does not speculate otherwise~\cite{ST_constantTime_Spec}. 
The only thing the oracle decides here is the size of the speculation window $\omega$.

\subsubsection{Symbolic Semantics}\label{sec:v5-symb}
Just as before, the symbolic speculative semantics $\semra$ replaces  concrete configurations and the non-speculative semantics with symbolic configurations and the symbolic non-speculative semantics respectively. 
We remark that the program counter $\pc$ is \emph{always} concrete in the symbolic non-speculative semantics~\cite{spectector}.
As a result, the RSB only contains concrete values (and return addresses).
The behaviour $\behRa{p}$ of a program $p$ is the set of all traces starting from an initial state until termination using the reflexive-transitive closure of the symbolic semantics

\subsubsection{Different Behaviours of Empty and Full RSBs}\label{sec:v5-rsb}
Modern CPUs use different RSBs implementations that differ in the way they handle under- and overflow, i.e., when the RSB is empty or full \cite{ret2spec}. 
For example, cyclic RSB implementations overwrite old entries when the RSB is full.
Alternatively, CPUs can fallback to other predictors (like the indirect branch predictor) to predict return addresses whenever the RSB is empty.

In our model, the RSB is not cyclic and there is no speculation when the RSB is empty (\Cref{tr:v5-retE}).
\begin{center}
    \typerule{$\Rvr$:AM-Ret-Empty}
    {
    \instr{\pret} & \conctauarrow{\sigma}{\sigma'} \\
    }
    {
    \tup{p, \ctr, \sigma, \mathbb{\emptyset}, n + 1}
    \specarrowR{\tau} 
    \tup{p, \ctr, \sigma', \mathbb{\emptyset}, n}
    }{v5-retE}
\end{center}

We remark that extending $\semr$ to support different RSBs implementations  can be done with minimal effort.

%% file: src/composing_semantics.tex
\section{A Framework for Composing Speculative Semantics}\label{sec:frame}
The presented speculative semantics allow us to verify programs for violations of \SNI{} but they do not capture the vulnerability in \Cref{lst:v1-v4-combined}, as the traces of \Cref{ex:comp-sni-iso} show.
\begin{example}[\SNI{} for \Cref{lst:v1-v4-combined}, \coqed]\label{ex:comp-sni-iso}
The traces generated are:
\begin{align*}
\begin{split}
\tauStack_{\Bv}^1 = \tauStack_{\Bv}^2 := {}& \storeObs{p} \concat \storeObs{p} \concat 
\startObsB{0} \concat \loadObs{p} \\
&  \concat \loadObs{A + public} \concat \rollbackObsB{0} \concat \pcObs{9}
\end{split}\\
\begin{split}
\tauStack_{\Sv}^1 = \tauStack_{\Sv}^2 := {}& ... \concat \storeObs{p} \concat \startObsS{1} \concat \skipObs{1} \concat \pcObs{\perp}  \concat \\
& \rollbackObsS{1} \concat \pcObs{\perp}
\end{split}
\end{align*}
The program in \Cref{lst:v1-v4-combined} seems secure since there is no secret value leaked in the speculative transaction; thus the program satisfies \SNI{} for $\sems$ and $\semr$ in isolation.
As we show later, the program in \Cref{lst:v1-v4-combined} is not secure, but we need our combined semantics to point out this vulnerability, as we show in \Cref{sec:comp-14}.
\end{example}

The vulnerability only appears when the branch predictor (\Cref{sec:v1-am}) and the memory disambiguator (\Cref{sec:v4-am}) are used \emph{together}.
Intuitively, we know that the CPU uses all the predictors described here (and many others as well) at the same time. 
Thus, we should not only focus on these different attacks in \textit{isolation} but we need to look at their \textit{combinations} as well.
That is, we need is a way to compose the different semantics into new semantics that can detect these combined vulnerabilities.

This section presents a novel, general framework for composing two speculative semantics $x$ and $y$ to allow for speculation from both sources $x$ and $y$. 
The speculative semantics $x$ and $y$ are also called the \textit{source} semantics of the composition. 
Thus, this section first introduces the new composed semantics, which consists of the always-mispredict, the oracle and the symbolic semantics (\Cref{sec:comb-am}). 
Finally, the section define what it means to be a \textit{proper} composition by identifying key correctness properties (\Cref{sec:comb-correctness}). 
From this definition, we can derive corollaries like that \SNI{} of the combination implies \SNI{} of both of its parts.

\paragraph{New Notation}
The states $\Sigmaxy$, instances $\Phixy$ and the trace model $\Obsxy$ are defined as the union of the source parts.
Furthermore, we define a projection function $\specProjectxy{}$ and two projections $\specProjectxyx{}$ and $\specProjectxyy{}$ that return the first and second projection of the pair from $\specProjectxy{}$. These functions are lifted to states by applying them pointwise:
\begin{align*}
    &\Obsxy \eqdef\ \Obsx \cup \Obsy  
    &
    &\Phixy \eqdef\ \Phix \cup \Phiy
    &
     \Sigmaxy \eqdef\ \Sigmax \cup \Sigmay\\
    &\specProjectxy{} \colon \Phixy \mapsto (\Phix, \Phiy)
    &
    &\specProjectxyx{} \colon\ \Phixy \mapsto \Phix 
    &
    \specProjectxyy{} \colon\ \Phixy \mapsto \Phiy 
\end{align*}

For example, the $\PhiSR$ states resulting of the union of $\PhiS$ and $\PhiR$ states (from \Cref{sec:v4-am} and \Cref{sec:v5-am} respectively), is $\tup{p, \ctr, \sigma, \Rsb, n}$, as it contains all the common elements (the program $p$, the counter $\ctr$, the state $\sigma$, and the speculation count $n$), plus the return stack buffer $\Rsb$ that belongs to $\PhiR$ only.
Taking the $\specProjectSSR{\cdot}$ of a $\PhiSR$ state returns the $\PhiS$ subpart (i.e., all but the return stack buffer).

We overload $\specProjectxyx{}$ and $\specProjectxyy{}$ to also work on traces $\tauStack$.
The projection on traces deletes all speculative transactions (marked by $\startObsy{\id}$ and $\rollbackObsy{\id}$) that are not generated by the source semantics $x$. The definition of $\specProjectxyy{}$ is similar by replacing $x$ with $y$:
\begin{align*}
    \specProjectxyx{\empTr} =~ \empTr 
    \qquad\qquad
    \specProjectxyx{(\tau \cdot \tauStack)} =&~ \tau \cdot \specProjectxyx{(\tauStack)}
    \\
    \specProjectxyx{(\startObsy{\id} \cdot \cdots \rollbackObsy{\id} \cdot \tauStack)} =&~ \specProjectxyx{\tauStack} 
\end{align*}
We indicate source semantics for $x$ and $y$ as $\semx$ and $\semy$ respectively and use $\semxy$ to indicate their composed semantics.

\subsection{Combined Speculative Semantics}\label{sec:comb-am}\label{sec:expl-z}

The idea behind the combined semantics is to delegate back to the source semantics of $x$ and $y$ to allow for speculation on both sources. This, in turn, yields proof reuse and is encapsulated in the two core rules below:
\begin{center}
    \typerule {AM-x-step}
    {\specProjectxyx{\Phixy} \specarrowxZ{\tau} \specProjectxyx{\phiStackxy'}
    }
    {\Phixy
    \specarrowxyZ{\tau}
    \phiStackxy'
     }{comb-x-step-Z}
     \typerule {AM-y-step}
    {\specProjectxyy{\Phixy} \specarrowyZ{\tau} \specProjectxyy{\phiStackxy'}
    }
    {\Phixy
    \specarrowxyZ{\tau}
    \phiStackxy'
     }{comb-y-step-Z}
\end{center}

The semantics does a step by either delegating back to the $x$ source semantics (\Cref{tr:comb-x-step-Z}) or to the $y$ one (\Cref{tr:comb-y-step-Z}). 
Each rule relies on metaparameter $\Zxy$, which is a pair of two metaparameters $\Zxy \eqdef (\Zx, \Zy)$ --- one for $x$ and one for $y$. 
We overload the projections $\specProjectxyx{}$ and $\specProjectxyy{}$ to extract he corresponding metaparameter from $\Zxy$. 

The role of $\Z$ is central to making the composed semantics work as expected, so now we explain $\Z$ in detail.
If we were to remove metaparameter $\Z$ and combine $\semb$ and $\sems$ into $\sembs$, consider the execution of the $\jzC{}$ instruction in \Cref{line:v14branch} in \Cref{lst:v1-v4-combined}. 
Without $\Z$, $\sembs$ can use \Cref{tr:comb-x-step-Z} and delegate back to $\semb$ for $\jzC{}$ instructions, creating a new speculative transaction. 
However, $\sembs$ can also use \Cref{tr:comb-y-step-Z}, because $\jzC{}$ instructions are also handled by $\sems$. 
Unfortunately, this would lead to no speculative transaction because speculation happens only on $\storeC{}$ instructions.
Intuitively, $\sembs$ should delegate back to $\semb$, so \Cref{tr:comb-y-step-Z} should not be applicable.

With metaparameter $\Z$, we can instantiate it for $\sembs$ as $\Zbs = ([\storeC{}], [\jzC{}])$, so that its projections are $\Zb = [{\storeC{}}]$ and $\Zs = [{\jzC{}}]$. 
Now, $\sembs$ can only apply \Cref{tr:comb-x-step-Z} on the $\jzC{}$ of \Cref{line:v14branch}, because $\Zs$ ensures that $\sems$ can not execute $\jzC{}$ instructions, as depicted in the full rule for $\sems$ below (where we indicate the instructions derived from $\Z$ in blue).
\begin{center}
    \typerule {$\Svr$:AM-NoSpec}
    {p(\sigma(\pc)) \notin [{\pstore{x}{e};  \pjzb{x}{\lbl}}] & \conctauarrow{\sigma}{\sigma'} 
    }
    {\tup{p, \ctr, \sigma, n + 1}
    \specarrowS{\tau}
    \tup{p, \ctr, \sigma', n}
     }{v4-noBranch}
\end{center}

Having clarified the intuition behind the semantics, we can define the behaviour $\behxy{}$as the set of all traces generated from an initial state until termination using the reflexive-transitive closure of $\specarrowxy{\tau}$.

\subsubsection{Oracle Combination}\label{sec:comb-oracle}
Instead of using one oracle, the combination uses a pair of oracles, one from each source. When delegating back to either source, the correct oracle of the source is handed over to the source semantics.

\subsubsection{Symbolic Combination}\label{sec:comb-symb}
Instead of using the AM semantics for delegation, the combined symbolic semantics $\semxya$ uses the symbolic source semantics for delegation. Furthermore, the new notation (union, projections) is lifted to the symbolic combination to create the symbolic states $\Sigmaxya$. The behaviour $\behxya{p}$ of program $p$ is the set of all traces generated from an initial state until termination using the reflexive-transitive closure of the symbolic semantics.

\subsection{Properties of Composition}\label{sec:comb-correctness}

We now illustrate the benefits of our composition framework.
For this, we first introduce a notion of well-formed composition (\Cref{sec:comb:correctness:wellformedness}), which intuitively tells when a combined semantics  ``makes sense''.
Then, we show that for well-formed compositions, if the source semantics are \sss, then the combined semantics is also \sss (\Cref{sec:comb:correctness:preservation}).
Note that since we prove these properties in the framework, any well-formed composition is also \sss \emph{for free}.
This proof reuse and extensibility is the key advantage of our framework over having ad-hoc semantics combining multiple speculation sources, which requires one to manually prove the \sss results we instead obtain for free.

\subsubsection{Well-formed Compositions}\label{sec:comb:correctness:wellformedness}
The well-formedness conditions for the composition ensures that the delegation is done properly (\Cref{def:wellformed}), they are the \emph{minimal} set of assumptions that let us derive \sss of the combined semantics for free.
For this, the well-formedness conditions attest that (1) the composed semantics is deterministic, and (2) the composed semantics does \emph{not} hide observations produced by its components.

Point (3) is fairly technical, and to present it, we need to mention a technical detail: the state relation (denoted $\srelxy$ and defined in our technical report) between the AM states ($\Sigmaxy$) and the Oracle ones ($\Xxy$).
Intuitively, two states are related if they are the same or if one is waiting on a speculation of the other to end. %
Then, point (3) ensures that whenever we start from related states ($\Sigmaxy\srelxy\Xxy$) and we do one or more steps of the AM composed semantics ($\Sigmaxy\specarrowxy{\tauStack}^* \Sigmaxy'$), then we can \emph{always} find a related state ($\Sigmaxy'\srelxy\Xxy'$)  that is reachable by performing one or more steps of the composed oracle semantics ($\Xxy \SEspecarrowxy{\tauStack'}\!\!^*\ \Xxy'$).
This fact is used when proving that \SNI{} of a program under the composed AM semantics implies \SNI{} under the composed oracle semantics (point 1 of \Cref{def:sss}).
Thus, it is not important for the AM and the Oracle semantics to produce the same traces, just that the two AM traces and the two Oracle traces are pairwise equivalent -- which follows from the state relation.
 
\begin{definition}[Well-formed composition]\label{def:wellformed}
A composition $\semxy$ of two speculative semantics $\semx$ and $\semy$ is \textit{well-formed}, denoted with $\wfc{\semxy}$ if:
\begin{compactenum}
    \item \textit{(Confluence)} Whenever $\Sigmaxy \specarrowxy{\tau} \Sigmaxy'$ and $\Sigmaxy \specarrowxy{\tau} \Sigmaxy''$, then  $\Sigmaxy' = \Sigmaxy''$.
    \item \textit{(Projection preservation)} For all programs $p$, $\behx{p} = \specProjectxyx{\behxy{(p)}}$ and $\behy{p} = \specProjectxyy{\behxy{(p)}}$.
    \item \textit{(Relation preservation)} If $\Sigmaxy\srelxy\Xxy$ and $\Sigmaxy\specarrowxy{\tauStack}^* \Sigmaxy'$ then $\Xxy \SEspecarrowxy{\tauStack'}\!\!^*\ \Xxy'$ and $\Sigmaxy'\srelxy\Xxy'$.
\end{compactenum}
\end{definition}

Since the combined semantics delegates back to its sources, it becomes non-deterministic. 
Confluence is needed to ensure the non-determinism is not harmful. 
Consider the assignment in \Cref{line:v14as} in \Cref{lst:v1-v4-combined}. 
$\sembs$ can delegate to either $\semb$ or $\sems$ to reduce the assignment.
If the combined semantics is \emph{Confluent} then it does not matter which source rule executes the assignment in  \Cref{line:v14as} in \Cref{lst:v1-v4-combined}, the semantics reaches the same state either way.

The projection preservation condition, instead, ensures that the combined semantics is not hiding or forgetting traces of its sources.
Any observable emitted by a source semantics must be propagated to the combined one, this is also the reason why $\Obsxy$ is defined as the union of the source $\Obs$.

\subsubsection{Free Theorems}\label{sec:comb:correctness:preservation}
The key result of our framework is that well-formed compositions whose source are secure speculative semantics (\sss) are also \sss (\Cref{thm:comp-sss}).

\begin{theorem}[$\semxy$ is \sss]\label{thm:comp-sss}
if $\ssssem{\semx}$ and $\ssssem{\semy}$ and $\wfc{\semxy}$ then $\ssssem{\semxy}$.
\end{theorem}

As a corollary of \Cref{thm:comp-sss}, we obtain that the security of well-formed compositions is related to the security of their components (\Cref{thm:comb:sni-preservation}).
In particular, whenever a program is insecure w.r.t. one of the components, then it is insecure w.r.t. the composed semantics.
Dually, if a program is secure w.r.t. the composed semantics, then it is secure w.r.t. the single components.
Note, however, that there are programs that are secure for the single components but insecure w.r.t. the composed semantics like \Cref{lst:v1-v4-combined}.

\begin{theorem}[Combined \SNI{} Preservation]\label{thm:comb:sni-preservation}
If $\wfc{\semxy}$ and $\nsnix$ or $\nsniy$, then $\nsnixy$.

If $\wfc{\semxy}$ and $\snixy$, then $\snix$ and $\sniy$.
\end{theorem}

Our free theorems have an immediate practical impact on \tool.
Note in fact that (1) \tool{}'s security analysis relies on the symbolic semantics, (2) the source symbolic semantics $\semb$, $\sems$, and $\semr$ are \sss, (3) well-formed compositions are also \sss, and (4) the composition of $\semb$, $\sems$, and $\semr$ are well-formed.
So, the \tool{} security analysis equipped with any combination of the $\semb$, $\sems$, and $\semr$ produces sound results, i.e.,  whenever the tool proves that a program is leak-free then the program satisfies \SNI{}.
So, the next section describes all the compositions and proves they are well-formed, while the section thereafter describes their implementation in \tool.\textbf{}

\section{Instantiating our Framework}\label{sec:comb-in}
This section describes all possible combinations of the speculative semantics $\semb$, $\sems$, and $\semr$: $\semsr$ (\Cref{sec:comp-45}), $\sembr$ (\Cref{sec:comp-15}), $\sembs$ (\Cref{sec:comp-14}),  and $\sembsr$ (\Cref{sec:comp-145}).
For each of them, we overview how the combined always-mispredict semantics behave using concrete examples and we prove that the combined semantics is well-formed, i.e., it satisfies \Cref{def:wellformed}.

In the following, we describe in detail how the $\semsr$ semantics can be instantiated as part of our framework; the other combinations can be instantiated similarly and we only provide a higher-level description.
We remark that the traces associated with the code snippets in this section have been computed using our Coq executable composed semantics.
Due to the length of these traces, here we elide uninteresting observations. 

\subsection{\texorpdfstring{$\semsr$}{Spec-S+R} Composition }\label{sec:comp-45}
To combine semantics using our framework, we need to define the states, observations, and  metaparameter $\Zsr$ for the composed semantics $\semsr$.
The combined state $\SigmaSR$ is the union of the states $\SigmaS$ and $\SigmaR$; thus it contains the RSB $\Rsb$ as well. 
\begin{gather*}
\begin{aligned}
\ti{Spec. States } \SigmaSR  \bnfdef&\ \phiStackSR
&
\ti{Spec. Instance } \PhiSR \bnfdef&\ \tup{p, \ctr, \sigma, \Rsb, n}
\end{aligned}
\end{gather*}
The union $\ObsSR{}$ of the trace models $\ObsS{}$ and $\ObsR{}$ is defined as:
\begin{gather*}
\begin{aligned}
\ObsSR{} \bnfdef&\ \startObsS{n} \mid \startObsR{n} \mid \rollbackObsS{n} \mid \rollbackObsR{n} \mid \skipObs{n} \mid ... 
\end{aligned}
\end{gather*}

To define the metaparameter  $\Zsr$, we need to identify, for each component semantics, the instructions that are related with speculative execution.
For $\sems$, the only instruction associated with speculative execution is $\storeC{}$, since the semantics can only speculative bypass stores.
For $\semr$, even though the semantics speculates only over $\retC{}$ instructions,  $\callC{}$ instructions also affect speculative execution since $\semr$ pushes  return addresses onto the $\Rsb$ when executing $\callC$. 
Therefore, we set the metaparameter $\Zsr$ to $([\callC{}, \retC{}], [\storeC{}])$.
This ensures that in the combined semantics $\semsr$, $\storeC{}$ instructions are only executed by delegating back to $\semsZ{\callC{}, \retC{}}$ and similarly, $\callC{}$ and $\retC{}$ instructions are only executed by delegating back to $\semrZ{\storeC{}}$.

\Cref{thm:v45-goodcomp} states that combining $\sems$ and $\semr$ in the way described above results in a well-formed composition.
Given that $\sems$ and $\semr$ are both \sss (\Cref{thm:v4-sss} and \Cref{thm:v5-sss}), we can derive ``for free'' that $\semsr$ is also \sss by applying \Cref{thm:comp-sss}.

\begin{theorem}[Well-formed Composition $\semsr$]\label{thm:v45-goodcomp}
$\wfc{\semsr}$
\end{theorem}

Let us now informally argue why \Cref{thm:v45-goodcomp} holds.
Intuitively, Confluence follows from the fact that both $\sems$ and $\semr$ delegate back to the non-speculative semantics for instructions not related to speculation. 
Since instructions related to speculation are already restricted by the metaparameter $\Zsr$, we only need to show \textit{Confluence} for those instructions not related to speculation.
Because the non-speculative semantics is deterministic, we can derive \textit{Confluence} for instructions not related to speculation.

For the Projection preservation, the composed semantics allows nesting of \textit{different} speculative transactions i.e., the nesting of speculative transactions for $\storeC{}$ and $\retC{}$ instructions. Despite the nesting, these transactions are still fully explored --- similar to what the corresponding source semantics would do.

\Cref{lst:v4-v5-comb} presents a program that contains a leak that can be detected only by $\semsr$ but not by its components $\sems$ and $\semr$.

\begin{lstlisting}[style=MUASMstyle, caption={$\semsr$ example}, label={lst:v4-v5-comb}, escapechar=|]
Manip_Stack:
    sp |$\xleftarrow{}$| sp + 8 |\label{line:v45-add}|
    ret
Speculate:
    call Manip_Stack |\label{line:v45-start-manip}|
    store secret, p |\label{line:v45x}|
    store pub, p |\label{line:v45s}|
    load eax, p |\label{line:v45l}|
    load edi, eax |\label{line:v45l2}|
    ret
Main:
    call Speculate |\label{line:v45start}|
    skip |\label{line:v45end}|
\end{lstlisting}

In \Cref{lst:v4-v5-comb}, execution starts on \Cref{line:v45start} by calling the function $\speculate$ and it continues at \Cref{line:v45-start-manip}. 
Next, the function $\manipStack{}$ is called and the stack pointer $\spR$ is incremented (\Cref{line:v45-add}).
This modifies the return address of the function $\manipStack{}$ to now point to \Cref{line:v45end} (the return address of the $\callC{}$ to $\speculate$).
Under $\semr$, mispredicting the return address of $\manipStack{}$ using the RSB leads continuing the execution at \Cref{line:v45x}. 
However, the $\storeC{}$ instructions in \Cref{line:v45s} overwrites the secret value stored in \Cref{line:v45x}.
Then, the $\loadC{}$ instructions in \Cref{line:v45l} and \Cref{line:v45l2} emit only public values. 
As a result, no secret is leaked and speculation ends.
Similarly, under $\sems$, speculation over store bypasses has no effect in  \Cref{lst:v4-v5-comb} because the $\storeC{}$ instruction in \Cref{line:v45x} is never reached and function $\manipStack{}$ returns to \Cref{line:v45end}. 
Therefore, the leak in missed under $\sems$ and $\semr$, i.e., $\text{\Cref{lst:v4-v5-comb}} \vdash_{\Sv} \text{\SNI}$ and $\text{\Cref{lst:v4-v5-comb}} \vdash_{ \Rv} \text{\SNI}$.

However, following the combined semantics $\semsr$, the $\storeC{}$ instruction on \Cref{line:v45s} is now skipped and when returning from function $\manipStack{}$ execution speculatively continues from \Cref{line:v45l}. 
Now, the $\loadC{}$ instructions are executed and the secret is leaked, as shown in the traces below.
Since $\mi{secret}$ is a high value, we can find two low-equivalent configurations $\sigma^1, \sigma^2$ that differ in the value of $\mi{secret}$.
This means we can find two traces (\coqed) that differs in the observation $\loadObs{secret}$ (highlighted in gray).
Thus, the program is not secure under the combined semantics i.e., $\text{\Cref{lst:v4-v5-comb}} \nvdash_{\Sv + \Rv} \text{\SNI}$.
\begin{align*}
\tau_{\Sv + \Rv}^2 \neq \tau_{\Sv + \Rv}^1 \isdef &\  
    \callObs{\mathit{Speculate}} \cdots \startObsR{0} \cdots \startObsS{1} \cdots \rollbackObsS{1} 
    \\
    \cdots&\ 
    \startObsS{2} \concat \skipObs{7} \concat \loadObs{p} \concat  \fcolorbox{lightgray}{lightgray}{$\loadObs{secret}$} 
    \cdots
\end{align*}
 
\newcommand\tikzmarkT[1]{%
  \tikz[remember picture,overlay]\node (#1) {};%
} 

The relation between the source semantics and their composition is visualised in \Cref{fig:comb-sem}, which shows the insecure programs (w.r.t \SNI{}) detected under the  different semantics.
The combined semantics encompasses all vulnerable programs of $\sems$ and $\semr$ \textit{and} additional programs like \Cref{lst:v4-v5-comb}.
These additional programs are the reason why the semantics $\semsr$ is `stronger than the sum of its parts' $\sems$ and $\semr$.
\begin{figure}[!ht]
    \centering
\begin{tikzpicture}[remember picture]

\tikzset{
 dot/.style = {circle, fill=black, minimum size=3pt,
               inner sep=0pt, outer sep=0pt},
 }

\draw [black, dashed, line width=0.5mm] plot [smooth cycle] coordinates {(-2,0) (-1,1) (1,1.5) (3,1.5) (6,0) (2,-1)};

\draw [\scol] plot [smooth cycle] coordinates {(-1,0) (1,0.5) (2,0.5) (2.5,0) (2,-0.5) };

\draw [\rcol] plot [smooth cycle] coordinates {(2,0.8) (4,0.5) (5,0.3) (5,0) (4,-0.5) };

\node[draw] at (3, 1.9)  {$\semsr$};

\node[draw, \scol] at (-0.6, 0.7)    {$\sems$};
\node[draw, \rcol] at (3.3, 1.0)    {$\semr$};

\node[dot,\scol] at (1,0) (V4E) {}; %
\node[dot, \rcol] at (4,0) (V5E) {}; %
\node[dot,black] at (1.5,1.1) (V45E) {}; %

\node[draw] at (-0.6, 1.7)   (V4S) {\Cref{lst:v4-vanilla}};
\node[draw] at (5.5, 1.4)    (V5S){\Cref{lst:v5-example}};
\node[draw] at (1.5, 2.1)    (V45S){\Cref{lst:v4-v5-comb}};

 \draw[->, overlay,\scol, thick, dotted,>=latex,] (V4S) -- (V4E);
 \draw[->, overlay, \rcol, thick, dotted,>=latex,] (V5S) -- (V5E);
 
 \draw[->, overlay,black, thick, dotted,>=latex,] (V45S) -- (V45E);

\end{tikzpicture}
\caption{Relating $\sems$, $\semr$ and $\semsr$ wrt \SNI{}.}
\label{fig:comb-sem}
\end{figure}

\subsection{\texorpdfstring{$\sembr$}{Spec-B+R} Composition }\label{sec:comp-15}
In this combination,  the instructions that influence speculative execution are $\callC{}$ and $\retC{}$ ($\semr$) and $\jzC{}$ ($\semb$). 
Thus, we set $\Zbr = ([\callC{}, \retC{}], [\jzC{}])$ to account for this, and to allow speculation from both sources.

\Cref{thm:v15-goodcomp} states that $\sembr$ is a well-formed composition.
As before, this allows us to derive ``for free'' that $\sembr$ is \sss by applying \Cref{thm:comp-sss}.

\begin{theorem}[Well-formed Composition $\sembr$]\label{thm:v15-goodcomp}
$\wfc{\sembr}$
\end{theorem}

\begin{lstlisting}[basicstyle=\small,style=MUASMstyle, caption={$\sembr$ example}, label=lst:v1-v5-comb, escapechar=|]
Manip_Stack:
    sp <- sp + 8 |\label{line:v15-add}|
    ret  |\label{line:v15-retM}|
Speculate:
    call Manip_Stack |\label{line:v15-start-manip}|
    x <- 0  |\label{line:v15x}|
    beqz x, L2 |\label{line:v15branch}|
    load eax, secret |\label{line:v15load}|
L2:
    ret |\label{line:v15ret}|
Main:
    call Speculate |\label{line:v15start}|
    skip  |\label{line:v15end}|
\end{lstlisting}
\Cref{lst:v1-v5-comb} presents a leak that can be detected only under $\sembr$.
The execution proceeds similarly to \Cref{lst:v4-v5-comb} until the $\retC{}$ instruction in \Cref{line:v15-retM} is reached.
Under $\semr$, mispredicting the return address leads to function $\manipStack{}$ returning to \Cref{line:v15x}. 
However, the $\jzC{}$ instructions in \Cref{line:v15branch} jumps to \Cref{line:v15ret} (since $x$ is 0) and speculation ends without leaking.
Under $\semb$, the branch instruction in \Cref{line:v15branch} is never executed and the function $\manipStack{}$ returns to \Cref{line:v15end} without leaking.
Hence, \Cref{lst:v1-v5-comb} is secure (i.e., it satisfies \SNI) when considering $\semr$ and $\semb$ in isolation.

Under the combined semantics $\sembr$, function $\manipStack$ returns to \Cref{line:v15x} and the $\jzC{}$ instruction is then mispredicted. 
This leads to executing the $\loadC{}$ instructions in \Cref{line:v15load}, which leaks secret information.
The resulting traces (\coqed) are given below, where we highlight the secret-dependent observations.
Given the length of the trace, we carve out only the most relevant parts, i.e., that both kinds of speculations need to have $\startObsKywd{}$ed in order for the leak to appear.
\begin{align*}
\tau_{\Bv + \Rv} ^2 \neq \tau_{\Bv + \Rv}^1 \isdef&\ 
\callObs{Speculate} \cdots \startObsR{0} \cdots \startObsB{1} 
    \\
    \concat &\ \pcObs{8} \concat \fcolorbox{lightgray}{lightgray}{$\loadObs{secret}$} \concat \rollbackObsB{1} \concat \rollbackObsR{0}
\end{align*}
Again, the two traces that differ in the observation in the grey box and we have $\text{\Cref{lst:v1-v5-comb}} \nvdash_{\Bv + \Rv} \text{\SNI}$.

\subsection{\texorpdfstring{$\sembs$}{Spec-B+S} Composition }\label{sec:comp-14}
In this combination,speculation happens on $\jzC{}$ instructions ($\semb$) and on $\storeC{}$ instructions ($\sems$). 
Thus, we set $\Zbs = ([\storeC{}], [\jzC{}])$. 
Therefore, in the combined semantics $\sembs$, $\jzC{}$ instructions are only executed by delegating back to $\sembZ{\storeC{}}$ and $\storeC{}$ instructions are only executed by delegating back to $\semsZ{\jzC{}}$.
This semantics is also a well-formed composition (\Cref{thm:v14-goodc}) and  \sss.
\begin{theorem}[Well-formed Composition $\sembs$]\label{thm:v14-goodc}
$\wfc{\sembs}$
\end{theorem}

\Cref{lst:v1-v4-combined} from  \Cref{sec:intro} contains a leak that can only be detected by $\sembs$ but not by its components.
The traces associated with the code (\coqed) are given below, where secret-dependent observations are highlighted in gray:
\begin{align*}
\tauStack_{\Bv + \Sv}^2 \neq \tauStack_{\Bv + \Sv}^1 \isdef&\  \cdots \startObsS{1} \concat \skipObs{1} \concat \cdots \startObsB{2} \concat \pcObs{5} \\
\concat  &\  \loadObs{p} \concat  \fcolorbox{lightgray}{lightgray}{$\loadObs{A + secret}$} \concat \rollbackObsB{2}  \concat \rollbackObsS{1} \cdots
\end{align*}
Thus, the program is not secure under $\sembs$, i.e., $\text{\Cref{lst:v1-v4-combined}} \nvdash_{\Bv + \Sv} \text{\SNI}$.

\subsection{\texorpdfstring{$\sembsr$}{Spec-B+S+R} Composition }\label{sec:comp-145}
We conclude this section by combining all three semantics $\semb$, $\sems$, and $\semr$.
Our framework (\Cref{sec:frame}) allows only combining a pair of source semantics into a combined one.
For simplicity, we present $\sembsr$ as a direct combination of the three source semantics (technically, we obtain $\sembsr$ by combining $\sembs$ with $\semr$).
The metaparameter $\Zbsr$ (which we represent as a triple of values) is $([\callC{}, \retC{}, \storeC{}], [\callC{}, \retC{}, \jzC{}], [\jzC{}, \storeC{}])$. 
As a result, the combined semantics $\sembsr$ can only delegate to the corresponding speculative semantics for the appropriate speculation sources.

As before, we can prove that $\sembsr$ is a well-formed composition (\Cref{thm:v145-goodc}) and we get that $\sembsr$ is \sss by applying \Cref{thm:comp-sss}.

\begin{theorem}[Well-formed Composition $\sembsr$]\label{thm:v145-goodc}
$\wfc{\sembsr}$
\end{theorem}

\begin{lstlisting}[style=MUASMstyle, caption={$\sembsr$ example}, label=lst:v1-v4-v5-comb, escapechar=|]
Manip_Stack:
    sp <- sp + 8
    ret
Speculate:
    call Manip_Stack
    x <- 0 |\label{line:v145r}| 
    beqz x, L2 |\label{line:v145branch}|
    load eax, p
    load edi, eax
L2:
    ret
Main:
    store secret, p 
    store pub, p |\label{line:v145s}|
    call Speculate
\end{lstlisting}
\Cref{lst:v1-v4-v5-comb} depicts a leaky program that can be detected only under $\sembsr$, since the program satisfies \SNI{} under $\semb$, $\sems$ and $\semr$. 
Under $\sembsr$, the $\storeC{}$ instruction in \Cref{line:v145s} is bypassed
Therefore, when returning from $\manipStack$, the program mispredicts the return address and speculatively returns to \Cref{line:v145r}. 
Here, the $\jzC{}$ instruction in \Cref{line:v145branch} is mispredicted and the $\loadC{}$ instructions are executed, which now leaks the secret value.

The resulting traces (\coqed) are given below:
\begin{align*}
\tau_{\Bv + \Sv + \Rv}^2 \neq \tau_{\Bv + \Sv + \Rv}^1 \isdef&\
    \cdots \startObsS{1} \concat \skipObs{14} \concat \callObs{\mathit{Speculate}} \cdots\\
    \concat &\ \startObsR{2} \concat \retObs{6} \concat \startObsB{3} \concat \pcObs{8} \concat \loadObs{p} \\
    \concat &\ \fcolorbox{lightgray}{lightgray}{$\loadObs{secret}$} \concat \rollbackObsB{3} \concat \rollbackObsR{2} \concat \concat \rollbackObsS{1} \cdots
\end{align*}
Thus, the program is not secure i.e., $\text{\Cref{lst:v1-v4-v5-comb}} \nvdash_{\Bv + \Sv + \Rv} \text{\SNI}$.

%% file: src/implementation.tex
\section{Implementation and Evaluation}\label{sec:impl-b}

\newcommand{\nrExamples}{49}

This section describes how our combined semantics can be used to detect leaks introduced by the interaction of multiple speculation mechanisms.
For this, we extended \tool{}, a symbolic analysis tool for speculative leaks, with the semantics for $\sems$, $\semr$ and for all the combinations presented in \Cref{sec:comb-in} (\Cref{sec:impl}).
Using \tool{}, we analyze a corpus of \nrExamples{} microbenchmarks containing speculative leaks generated by different speculation mechanisms (\Cref{sec:bench}).
With these experiments, we aim to show that (1) our $\sems$ and $\semr$ speculative semantics can correctly identify speculative leaks associated with speculation over store-bypasses and return instructions, and (2) our combined semantics can detect novel leaks that are otherwise undetectable when considering single speculation mechanisms in isolation.

\begin{table*}

\begin{subtable}[t]{.27\linewidth}
\vspace{0pt}
\begin{tabular}{llccc}
\toprule
\multirow{2}{*}{Test case} &  & \multicolumn{2}{c}{$\sems$} \\  \cline{3-4}
    &  &      None & Fence \\
\midrule
case01 &(\minusR)  & $\N$ & $\P$ \\
case02 &(\minusR) & $\N$ & $\P$ \\
case03 & (\plusG) & $\P$ & $\P$ \\
case04 &  (\minusR)& $\N$ & $\P$ \\
case05 &(\minusR) & $\N$ & $\P$ \\
case06 &(\minusR) & $\N$ & $\P$\\
case07 &(\minusR) & $\N$ & $\P$ \\
case08 &(\minusR) & $\N$ & $\P$ \\
case09&     (\plusG) & $\P$ & $\P$ \\
case10      &(\minusR) & $\N$ & $\P$ \\
case11      &(\minusR) & $\N$ & $\P$ \\
case12      &(\plusG) & $\P$ & $\P$ \\
case13 &   (\minusR) & $\N$ & $\P$ \\
\midrule

\end{tabular}
\subcaption{Results for the Spectre-STL programs under the $\sems$ semantics against unpatched programs (column ``None'') and programs patched with \texttt{lfence} (column ``Fence'')}
\label{t:table-v4}
\end{subtable}
\hspace{3em}
\begin{subtable}[t]{.45\linewidth}

\vspace{0pt}
\begin{tabular}{llcccc}
\toprule
\multirow{2}{*}{Test case}  &  &  \multicolumn{3}{c}{$\semr$} \\ \cline{3-5}
&  & None & Fence & Retpoline \\
\midrule
$ret2spec\_c\_d$ &(\minusR)  & $\N$ & $\P$ & $\P$ \\
$ca\_ip$& (\minusR)  & $\N$ & $\P$ & $\P$\\
$ca\_oop$& (\minusR)  & $\N$ & $\P$ & $\P$\\
$sa\_ip$ &(\minusR)  & $\N$ & $\P$ & $\P$\\
$sa\_oop$ & (\minusR)  & $\N$ & $\P$ & $\P$\\
\bottomrule
\end{tabular}
\subcaption{Results for the Spectre-RSB programs under the $\semr$ semantics against unpatched programs (column ``None''), programs patched with \texttt{lfence} (column ``Fence''), and programs patched with retpoline (column ``Retpoline'')}
\label{t:table-v5}
\vspace{2em}
\begin{tabular}{llccccccc}
\toprule
Test case &   & $\semb$ & $\sems$ & $\semr$ & $\sembs$ & $\semsr$ & $\sembr$ & $\sembsr$ \\
\midrule
\cref{lst:v1-v4-combined}& (\minusR) & $\P$ & $\P$ & $\P$ & $\N$ & $\P$ & $\P$ & $\N$ \\
\cref{lst:v1-v5-comb} &(\minusR) & $\P$ & $\P$ & $\P$  & $\P$ & $\N$ & $\P$ & $\N$\\
\cref{lst:v4-v5-comb} &(\minusR) & $\P$ & $\P$ & $\P$  & $\P$ & $\P$ & $\N$ & $\N$\\
\cref{lst:v1-v4-v5-comb}& (\minusR) & $\P$ & $\P$ & $\P$  & $\P$ & $\P$ & $\P$ & $\N$ \\
\cref{lst:v1-v4-combined} Fence& (\plusG) & $\P$ & $\P$ & $\P$ & $\P$ & $\P$ & $\P$ & $\P$ \\
\cref{lst:v1-v5-comb} Fence &(\plusG) & $\P$ & $\P$ & $\P$  & $\P$ & $\P$ & $\P$ & $\P$\\
\cref{lst:v4-v5-comb} Fence&(\plusG) & $\P$ & $\P$ & $\P$  & $\P$ & $\P$ & $\P$ & $\P$\\
\cref{lst:v1-v4-v5-comb} Fence& (\plusG) & $\P$ & $\P$ & $\P$  & $\P$ & $\P$ & $\P$ & $\P$ \\
\bottomrule

\end{tabular}
\subcaption{Results for the Spectre-Comb programs where ``listing $x$ Fence'' denotes the patched version (using \texttt{lfence}) of ``listing $x$''}
\label{t:table-comb}
\end{subtable}

\caption{Result of the analysis of our benchmarks. 
For each program,  $\N$ denotes that our tool finds a violation of SNI w.r.t to the corresponding semantics, whereas $\P$ denotes that our tool proves the program secure under the semantics.
Next to each program, we report whether the program contains a speculative leak (\plusG) or not (\minusR) in its unpatched version.
}
\label{tab:evaluation}
\end{table*}

\subsection{Implementation}\label{sec:impl}
We implemented all our semantics (the symbolic versions of $\sems$ and $\semr$ plus all the compositions from  \Cref{sec:comb-in}) as an extension of \tool \cite{spectector}. \tool uses symbolic execution together with self-composition \cite{R_self_comp} and a SMT solver to check for \SNI{} w.r.t. $\semb$. 
Due to this setup, we inherit limitations of \tool as well, i.e., path explosion because of symbolic execution and limitations in the translation from x86 to \muasm{}.

\subsection{Experiments}\label{sec:bench}

\tightpar{Benchmarks}
We analyze \nrExamples{} snippets of code containing leaks resulting from speculation over branch, store/load, and return instructions (and their combinations):
\begin{asparaitem}
    \item \textbf{Spectre-STL:} 13 programs are variants of the Spectre-STL vulnerability. 
    They exploit speculation over memory disambiguation, and they have been used as benchmarks in prior work~\cite{ST_binsec,cats}.
    For each program, we also analyze a patched version where a manually inserted \textsc{lfence} instruction stops speculation over store-bypasses and prevents the leak.
    
    \item  \textbf{Spectre-RSB:} 5 programs are variants of the Spectre-RSB vulnerability. 
    They exploit speculation over return instructions, and they are obtained from the \texttt{safeside}~\cite{safeside}\footnote{Out of the three Spectre-RSB examples from~\cite{safeside}, we analyze the only one that works against an acyclic RSB like the one supported by $\semr$.}  and \texttt{transientfail}~\cite{transientfail}\footnote{Programs \textit{ca\_ip}, \textit{ca\_oop}, and \textit{sa\_ip} from \texttt{transientfail} rely on concurrent execution.
Since \tool{} does not support concurrency, we hardcode the worst-case interleaving in terms of speculative leakage in our benchmark. } projects.
    For each program, we also analyze manually patched versions obtained by (1) inserting \textsc{lfence}s after each call instruction (i.e., at the instruction address where $\pret$ speculatively returns), and (2) using the  \texttt{retpoline} defense~\cite{ret2spec}.
    
    \item \textbf{Spectre-Comb:} 4 programs contain leaks that arise from combining speculation mechanisms.
    These are the programs depicted in \cref{lst:v1-v4-combined}, \cref{lst:v1-v5-comb}, \cref{lst:v4-v5-comb}, and \cref{lst:v1-v4-v5-comb} and discussed in \Cref{sec:comb-in}.
    For each program, we also analyze a manually patched version where \texttt{lfence} instructions prevent the speculative leaks.
\end{asparaitem}

\tightpar{Experimental setup}
The benchmarks for \textbf{Spectre-STL} and \textbf{Spectre-RSB} are implemented in C and compiled with \gcc{} 11.1.0 and we manually inserted \texttt{lfence}/\texttt{retpoline} countermeasures in the patched versions.
The benchmarks for \textbf{Spectre-Comb} are directly formalized in \muasm{}.
We run all our experiments on a  laptop with a Dual Core Intel Core i5-7200U CPU and 8GB of RAM.

\tightpar{Spectre-STL}
\Cref{t:table-v4} reports the results of analysing the programs in the \textbf{Spectre-STL} benchmark\footnote{We had to slightly modify programs 02, 05, and 06 due to limitations of \tool{}'s x86 front-end when dealing with global values (programs 05 and 06) and 32-bit addressing (program 02). We had to limit the speculation window, due to vanilla \tool{}'s limitations in symbolic execution, when analyzing program 09, which contains a loop.}.
Using the $\sems$ semantics, \tool{} successfully detected leaks (i.e., violations of \SNI) in all unpatched programs, except programs 03, 09, and 12 which do not contain speculative leaks (consistently with other analysis results~\cite{ST_binsec,cats}).
Observe that Binsec/Haunted~\cite{ST_binsec} flags program 13 as secure since the program can \emph{only} speculatively leak initial values from the stack, which Binsec/Haunted treats as public by default~\cite{hauntedBugReport}.
Since we assume initial memory values to be secret (like~\citet{cats}), \tool{} correctly detected the leak in program 13.
\tool{} also successfully proved that all patched programs (where an \texttt{lfence} is added between $\storeC{}$ instructions) satisfy \SNI{} and are free of speculative leaks.

\tightpar{Spectre-RSB}
\Cref{t:table-v5} reports the analysis results on the \textbf{Spectre-RSB} programs.
Using $\semr$, \tool{} successfully detected leaks in all unpatched programs.
Moreover, \tool{} successfully proved that the patched programs where a \texttt{lfence} instruction is added after every $\callC{}$ satisfy \SNI{}, i.e., they are free of speculative leaks.
\tool{} also successfully proved secure the programs patched using the modified \texttt{retpoline} defense proposed by \citet{ret2spec}, which replaces return instructions with a construct that traps the speculation in an infinite loop until speculation ends.

\tightpar{Spectre-Comb}
\Cref{t:table-comb} reports the results of our analysis on the \textbf{Spectre-Comb} programs, which involve leaks arising from a combination of multiple speculation mechanisms.
\tool{} equipped with the single semantics $\semb$, $\sems$, and $\semr$ is not able to detect the speculative leaks in any of the 4 programs and, therefore, proves them secure.
This is expected since the programs contain leaks that arise from a combination of semantics.
\tool{} can successfully identify leaks in \cref{lst:v1-v4-combined}, \cref{lst:v1-v5-comb}, \cref{lst:v4-v5-comb} when using, respectively, the semantics $\sembs$, $\semsr$, and $\sembr$.
Each semantics, however, fail in detecting leaks in the other programs, and all of them fail in detecting a leak in \cref{lst:v1-v4-v5-comb} as expected.
Finally, \tool{} is able to successfully detect leaks in all programs when using the $\sembsr$ semantics that combines all speculation mechanisms studied in this paper.

Similarly to \textbf{Spectre-STL} and \textbf{Spectre-RSB}, we also analyzed programs that have been manually patched by inserting \texttt{lfence} statements (``\cref{lst:v1-v4-combined} Fence'', ``\cref{lst:v1-v5-comb} Fence'', ``\cref{lst:v4-v5-comb} Fence'', and ``\cref{lst:v1-v4-v5-comb} Fence'' in \Cref{t:table-comb}).
As before, \tool{} successfully prove the security of patched programs.
We remark that, even for leaks that arise from multiple speculation sources, it is often sufficient to insert a single \texttt{lfence} to secure the entire program.
For instance, it is sufficient to introduce a \texttt{lfence} after the $\jzC{}$ instruction in \Cref{lst:v1-v5-comb} to ensure that the program satisfies \SNI{} with respect to $\sembsr$.

%% file: src/related_work.tex
\section{Related Work}\label{sec:rw}

\tightpar{Speculative execution attacks}
After Spectre~\cite{spectre} has been disclosed to the public in 2018, researchers have identified many other speculative execution attacks~\cite{spectreRsb,ret2spec,S_smotherSpectre,S_trans_troj,barberis2022branch}.
These attacks differ in the exploited speculation sources~\cite{ret2spec, spectreRsb, S_specv4}, the covert channels~\cite{S_specPrime, S_netSpectre, S_lazyFP} used, or the target platforms~\cite{chen2018sgxpectre}. 
We refer the reader to~\citet{transientfail} for a survey of existing attacks.

\tightpar{Security conditions for speculative leaks}
Researchers have proposed many program-level properties that capture different flavors of security against speculative leaks.
These properties can be classified in three main groups~\cite{sok:spectre_defense}:
\begin{asparaenum}
\item Non-interference definitions ensure the security of speculative \emph{and} non-speculative instructions. 
For instance, speculative constant-time~\cite{ST_constantTime_Spec} (used also in \cite{ST_jasmin2,ST_blade,ST_binsec}) extends the constant-time security condition to account also for transient instructions. 

\item Relative non-interference definitions~\cite{spectector,ST_tpod,ST_inspectre,ST_specusym} ensure that transient instructions do not leak more information than what is leaked by non-transient instructions.
For instance, speculative non-interference~\cite{spectector}, which we build on, (used also in~\cite{ST_spectector2,S_sec_comp}) restricts the amount of information that can be leaked by speculatively executed instructions (without constraining what can be leaked non-speculatively).

\item Definitions that formalise security as a safety property~\cite{cats,S_sec_comp}, which often over-approximate the non-interference definitions from the categories above.
\end{asparaenum}

\tightpar{Operational semantics for speculative leaks}
In the last few years, there has been a growing interest in developing formal models and principled program analyses for detecting leaks caused by speculatively executed instructions.
We refer the reader to~\cite{sok:spectre_defense} for a comprehensive survey on the topic.
In the following, we discuss the approaches that are more relevant to our paper.

Our speculative semantics $\sems$ and $\semr$ capture the effects of transient instructions at a rather high-level, and they are inspired by the always-mispredict $\semb$ semantics from~\cite{spectector}.
In contrast, other approaches, which we overview next, consider more complex models that explicitly model microarchitectural components like multiple pipeline stages, caches, and branch predictors. 

For instance, KLEESpectre~\cite{kleeSpectre} and SpecuSym~\cite{ST_specusym} consider a semantics that explicitly model the cache, which enable reasoning about the cache content.
\citet{R_sem_model2} go a step further and model a multi-stage pipeline with explicit cache and branch predictor.
Their semantics can only model speculation over branch instructions since it lacks store-forwarding or RSB.

\citet{ST_constantTime_Spec}'s semantics model speculation over branch instructions, store-bypasses, and return instructions.
Differently from our high-level semantics, their 3-stage pipeline semantics explicitly models several microarchitectural components like a reorder buffer and an RSB.
Their tool uses symbolic execution to detects violations of speculative constant-time induced by speculation over branch instructions and store-bypasses.

Binsec/Haunted~\cite{ST_binsec} can also detect violations of speculative constant-time induced by speculation over store-bypasses and branch instructions.
For this, their semantics explicitly model the store buffer, which $\sems$ abstracts away.
\citet{ST_jasmin} extend the Jasmin~\cite{ST_jasmin2} cryptographic verification framework to reason about speculative constant-time and, again, it supports speculation over store-bypasses and branch instructions.

While several of the models describe above support multiple speculation mechanisms, these mechanisms are \emph{hard-coded} and none of the aforementioned approaches provide a composition framework like ours (or extensible ways of extending the main theoretical results to new mechanisms ``for free'').
Moreover, while we could have used other operational semantics, like the one in~\cite{ST_constantTime_Spec}, as a basis for our composition framework, this would have resulted in more difficult proofs (since semantics like the one in~\cite{ST_constantTime_Spec} are significantly more complex than ours).

\tightpar{Axiomatic semantics for speculative leaks}
A few approaches formalize the effects of speculatively executed instructions using axiomatic semantics inspired by work on weak memory models.
For instance, \citet{ST_abs_sem} and \citet{R_sem_model} capture the effects of speculation over branch instructions but they both lack program analyses. 

\citet{cats} illustrate how one can model leaks resulting from speculation over branch instructions and load/store instructions using the CAT modeling language for memory consistency, and they present a bounded model checking analysis for detecting speculative leaks.
Interestingly, they talk about composing several of theirs semantics \cite[IV E.]{cats}, which in theory should allow them to detect vulnerabilities like \Cref{lst:v1-v4-combined} (which we detect under $\sembs$).
Differently from our framework, however, they do not formally characterize compositions and, therefore, they cannot derive interesting results ``for free'' about the composed semantics (like we do in \Cref{thm:comp-sss}).
Moreover, even though they state that composability is a main advantage of axiomatic models, our framework (and tool implementation) clearly show that composability can be done with operational semantics as well.

%% file: src/future_work.tex
\section{Conclusion and Future Work}\label{sec:conc}
This paper presented new speculative semantics for speculation on store instructions and on return instructions. 
Furthermore, it defined a general framework to reason about the composition of different speculative semantics and instantiated the framework with our new speculative semantics $\sems$ and $\semr$ and the semantics by \citet{spectector}. 
Our framework yields security of the composed semantics (almost) for free, given the security of its parts.
All the new semantics are implemented as extension in the tool \tool and the tool correctly detects all vulnerabilities in existing as well as in novel benchmarks.

There are multiple directions for future works.
Our composition is restricted in the sense that the different speculation sources do not influence each other.
\citet{spectreRsb} argued that during a transient execution caused by branch misprediction, one can add entries to the RSB by transiently executing a call instruction. During the roll back, the RSB is not rolled back, which could lead to more speculation down the line.
However, this composition would not be $\wfc$ because this composition would not be related to its projection ( for a rather technical reason one can see in the technical report). In the future, we would like to allow these combinations as well without losing all the `free' theorems presented here.

\tightpar{\textit{Acknowledgments}}
This work was partially supported 
    by the Madrid regional government under the project S2018/TCS-4339 BLOQUES-CM,  
    by the Spanish Ministry of Science, Innovation, and University under the project RTI2018-102043-B-I00 SCUM, and 
    by a gift from Intel Corporation,
    the Italian Ministry of Education through funding for the Rita Levi Montalcini grant (call of 2019),
    by the German Ministry for Education and Research through funding for the project CISPA-Stanford Center for Cybersecurity (Funding number: 16KIS0761).